\documentstyle[aps,epsf,prl]{revtex}
\begin{document}
\twocolumn[ %typeset the title and abstract in one column
\preprint{\vbox{\hbox{UCSD/PTH 95--08} \hbox{CALT--68--2000}
\hbox{hep-ph/9506356} }}
\title{Chiral Perturbation Theory for Vector Mesons}
\author{Elizabeth Jenkins and Aneesh V. Manohar}
\address{Department of Physics, University of California at San Diego \\
9500 Gilman Drive, La Jolla, CA 92093-0319}
\author{Mark B. Wise}
\address{California Institute of Technology, Pasadena, CA 91125}
\date{June 1995} \maketitle \widetext
\vskip-2.0in
\rightline{\vbox{ \hbox{UCSD/PTH 95--08} \hbox{CALT--68--2000}
\hbox{hep-ph/9506356} }}
\vskip1.5in
\begin{abstract} The interactions of $\rho$, $K^*$, $\phi$ and $\omega$
vector-mesons with low-momentum $\pi$, $K$ and $\eta$ pseudoscalar mesons are
constrained by chiral symmetry.  We derive a heavy vector-meson chiral
Lagrangian in which the vector mesons are treated as heavy static matter
fields.  The unknown couplings of the chiral Lagrangian are further related
using the $1/N_c$ expansion.  Chiral perturbation theory is applied to the
vector-meson mass matrix.  At one-loop there are large corrections to the
individual vector meson masses, but the singlet-octet mixing angle remains
almost unchanged. The parity-violating $s$-wave $\phi \rightarrow \rho \pi$
weak decay amplitude is derived in the combined chiral and large $N_c$
limits. Rare $\phi$ decays provide a sensitive test of non-leptonic neutral
current structure.
\end{abstract}


] % end two column format

\narrowtext

An important application of chiral perturbation theory is to describe the
interactions of matter fields (such as nucleons~\cite{jm} or hadrons containing
a heavy quark~\cite{lots}) with low-momentum pseudo-Goldstone bosons --- the
pions, kaons and eta.  In this paper we use chiral perturbation theory to
describe the interactions of the $\rho$, $K^*$, $\phi$ and $\omega$ vector
mesons with low-momentum pseudo-Goldstone bosons. The results of this work are
relevant for experiments at the $\phi$ factory being built at
Frascati\cite{daphne}. We apply chiral perturbation theory to transitions of
the form $V\rightarrow V^\prime X$, where $V$ and $V^\prime$ are vector
mesons. The mass differences between the nine lowest-lying vector mesons are
small compared with the chiral symmetry breaking scale of $\sim 1$~GeV, so
chiral perturbation theory is applicable as a systematic expansion procedure
for such decays.  Chiral perturbation theory has previously been used to study
processes such as $\rho \rightarrow \pi \pi$ which do not have a vector meson
in the final state.  Decays such as $\rho \rightarrow \pi \pi$ do not have soft
pions in the final state, so the application of the chiral lagrangian to such
processes is not justified, and should be considered as a phenomenological
model.

The pseudo-Goldstone boson fields can be written as a $3\times 3$ special
unitary matrix
\begin{equation} \Sigma = \exp {2i\bbox{\Pi}\over f} \label{eq:1}
\end{equation}
where
\begin{equation}\label{eq:2}
\bbox{\Pi} = \left[
\begin{array}{ccc}
{\pi^0\over\sqrt{2}} + {\eta\over \sqrt{6}} & \pi^+ & K^+ \\ \pi^- & -{\pi^0
\over \sqrt{2}} + {\eta \over \sqrt{6}} & K^0 \\ K^- & \overline K^0 & - {2\eta
\over \sqrt{6}}
\end{array} \right].
\end{equation}
Under chiral $SU(3)_L\times SU(3)_R$, $\Sigma \rightarrow L\Sigma R^\dagger$,
where $L\in SU(3)_L$ and $R\in SU(3)_R$.  At leading order in chiral
perturbation theory, $f$ can be identified with the pion or kaon decay constant
$(f_\pi \simeq 132\ {\rm MeV}, f_K \simeq 160\ {\rm MeV})$.  It is convenient,
when describing the interactions of the pseudo-Goldstone bosons with other
fields to introduce
\begin{equation}
\xi = \exp {i\bbox{\Pi} \over f}= \sqrt{\Sigma}. \label{eq:4}
\end{equation}
Under chiral $SU(3)_L \times SU(3)_R$,
\begin{equation}
\xi \rightarrow L \xi U^\dagger = U \xi R^\dagger, \label{eq:5}
\end{equation}
where in general $U$ is a complicated function of $L, R$ and the meson fields
$\bbox{\Pi}$.  For transformations $V = L = R$ in the unbroken $SU(3)_V$
subgroup, $U = V$.

The vector meson fields are introduced as a $3 \times 3$ octet matrix
\begin{equation}\label{eq:6}
{\cal O}_\mu = \left[
\begin{array}{ccc}
{\rho^0_\mu \over \sqrt{2}} + {\phi_\mu^{(8)} \over \sqrt{6}} & \rho_\mu^+ &
K_\mu^{*+} \\ \rho_\mu^- & -{\rho_\mu^0 \over \sqrt{2}} + {\phi_\mu^{(8)} \over
\sqrt{6}} & K_\mu^{*0} \\ K_\mu^{*-} & \overline K_\mu^{*0} & -{2\phi_\mu^{(8)}
\over \sqrt{6}}
\end{array} \right],
\end{equation}
and as a singlet
\begin{equation}
S_\mu = \phi_\mu^{(0)}. \label{eq:7}
\end{equation}
Under chiral $SU(3)_L \times SU(3)_R$,
\begin{equation}
{\cal O}_\mu \rightarrow U {\cal O}_\mu U^\dagger,\qquad S_\mu \rightarrow
S_\mu,
\label{eq:8}
\end{equation}
and under charge conjugation,
\begin{equation}\label{eq:9}
C {\cal O}_\mu C^{-1} = - {\cal O}_\mu^T, \ CS_\mu C^{-1} = - S_\mu, \ C \xi
C^{-1} = \xi^T.
\end{equation}

We construct a chiral Lagrangian for vector mesons by treating the vector
mesons as heavy static fields \cite{eh,g} with fixed four-velocity $v^\mu$,
$v^2 = 1$. The three polarization states of vector mesons with velocity $v^\mu$
satisfy $v \cdot S = v \cdot {\cal O} = 0$.  The chiral Lagrange density which
describes the interactions of the vector mesons with the low-momentum $\pi$,
$K$ and $\eta$ mesons has the general structure
\begin{equation}\label{eq:10}
{\cal L} = {\cal L}_{\rm kin} + {\cal L}_{\rm int} + {\cal L}_{\rm mass}.
\end{equation}
At leading order in the derivative and quark mass expansions,
\begin{eqnarray}
{\cal L}_{\rm kin} &=& - i\, S_\mu^\dagger (v \cdot \partial) S^\mu - i\, {\rm
Tr}\, {\cal O}_\mu^\dagger (v \cdot {\cal D}) {\cal O}^\mu,\label{kin}
\end{eqnarray}
and
\begin{eqnarray}
{\cal L}_{\rm int} &=& i g_1 \, S_\mu^\dagger\, {\rm Tr}\, \left({\cal O}_\nu
A_\lambda \right) v_\sigma \epsilon^{\mu \nu \lambda \sigma} + h.c. \nonumber\\
&&+ i g_2 \,{\rm Tr} \left( \left\{ {\cal O}_\mu^\dagger, {\cal O}_\nu\right\}
A_\lambda \right) v_\sigma \epsilon^{\mu \nu \lambda \sigma},
\label{eq:12}
\end{eqnarray}
where
\begin{equation}
{\cal D}^\nu {\cal O}^\mu = \partial^\nu {\cal O}^\mu + \left[ V^\nu, {\cal
O}^\mu \right],\label{cov}
\end{equation}
and
\begin{equation}V^\mu = \frac 1 2
\left(\xi \partial^\mu \xi^\dagger +\xi^\dagger \partial^\mu \xi \right) ,\
A^\mu = \frac i 2 \left(\xi \partial^\mu \xi^\dagger -\xi^\dagger \partial^\mu
\xi \right) .
\end{equation}
The terms in ${\cal L}_{\rm kin}$ appear with minus signs because the
polarization vector is spacelike.  Charge conjugation invariance requires that
the product of ${\cal O}_\mu^\dagger$ and ${\cal O}_\nu$ in the second term of
Eq.~(\ref{eq:12}) be an anticommutator.\footnote{It is important to remember
that in the matrix ${\cal O}_\mu^\dagger$ the field $\rho_\mu^{-\dagger}$ is
not equal to $\rho_\mu^+$, etc.  In heavy vector meson chiral perturbation
theory, $\rho_\mu^+$ destroys a $\rho^+$ but it does not create the
corresponding antiparticle.  A separate field $\rho_\mu^{-\dagger}$ is
introduced to create a $\rho^-$.}  Finally, to linear order in the quark mass
expansion,
\begin{eqnarray}
{\cal L}_{\rm mass} &=& \mu_0\, S_\mu^\dagger S^\mu + \mu_8\, {\rm Tr}\, {\cal
O}_\mu^\dagger {\cal O}^\mu \nonumber\\ &&+ \lambda_1\, \left({\rm Tr}\, \left(
{\cal O}_\mu^\dagger {\cal M}_\xi \right) S^\mu + h.c \right) \nonumber\\ &&+
\lambda_2\, {\rm Tr}\, \left( \left\{ {\cal O}_\mu^\dagger, {\cal O}^\mu
\right\} {\cal M}_\xi \right)
\label{eq:13} \\
&&+ \sigma_0\, {\rm Tr}\, {\cal M}_\xi \ S_\mu^\dagger S^\mu + \sigma_8\, {\rm
Tr}\, {\cal M}_\xi \ {\rm Tr}\, {\cal O}_\mu^\dagger {\cal O}^\mu , \nonumber
\end{eqnarray}
where $\cal M$ is the quark mass matrix $ {\cal M} = {\rm diag} \left(m_u, m_d,
m_s\right)$, and
\begin{equation}
{\cal M}_\xi = {1 \over 2} \left( \xi {\cal M} \xi + \xi^\dagger {\cal M}
\xi^\dagger\right).
\end{equation}
Note that the fields $S$ and ${\cal O}$ appearing in
Eqs.~(\ref{kin})--(\ref{cov}) are understood to be velocity-dependent fields
which are rescaled by a common phase factor (either $e^{-i\mu_{0} v \cdot x}$
or $e^{-i\mu_{8} v \cdot x}$).\footnote{The velocity-dependent vector meson
fields are related to the vector meson fields by $\phi^\mu_v = \sqrt{2 m}\,
e^{i m v \cdot x}\, \phi^\mu$, and have dimension 3/2.}  This rescaling removes
either $\mu_0$ or $\mu_8$ from Eq.~(\ref{eq:13}), so only the singlet-octet
mass difference $\Delta \mu \equiv \mu_0 - \mu_8$ is relevant.
Phenomenologically, the parameter $\Delta \mu < 200$~MeV is comparable to mass
splittings of order $m_s$, so in our power counting we treat $\Delta\mu$ as a
quantity of order $m_q$.  $\Delta\mu$ is of order $1/N_c$, and so vanishes in
the large $N_c$ limit.

We begin by considering the spectrum of vector mesons produced at leading order
in chiral perturbation theory. The analysis is identical to the well-known
$SU(3)$ analysis \cite{Kok}.  Neglecting isospin breaking due to the up and
down quark mass difference, i.e. $m_u= m_d = \hat m$, we find that
\begin{mathletters}\label{eq:16}
\begin{equation}
m_\rho = \bar \mu_8 + 2 \lambda_2 \hat m,\ \ m_{K^{*}} = \bar \mu_8 + \lambda_2
\left( \hat m + m_s \right),
\end{equation}
and the $\phi^{(0)} - \phi^{(8)}$ mass matrix is
\begin{equation}
M^{(08)} = \left[
\begin{array}{cc}
\bar \mu_0 & -{2\over\sqrt{6}} \lambda_1 \left( m_s - \hat m \right) \\
-{2\over\sqrt{6}} \lambda_1 \left( m_s - \hat m \right) & \bar \mu_8+{2\over 3}
\lambda_2 \left( \hat m + 2 m_s \right)
\end{array} \right],
\end{equation}
\end{mathletters}
where
\begin{equation}
\bar \mu_0 = \mu_0 + \sigma_0 {\rm Tr} {\cal M},\ \ \bar \mu_8 = \mu_8 +
\sigma_8
{\rm Tr} {\cal M} .
\end{equation}
Using Eqs.~(\ref{eq:16}a) and (\ref{eq:16}b), it is possible to express the
elements of $M^{(08)}$ in terms of the measured vector-meson masses (up to a
sign ambiguity for $M_{12}^{(08)}$)
\begin{mathletters}\label{eq:17}
\begin{eqnarray}
M_{11}^{(08)} &=& m_\omega + m_\phi - {4\over 3} m_{K^{*}} + {1\over 3} m_\rho
\\ M_{22}^{(08)} &=& {4\over 3} m_{K^{*}} - {1\over 3} m_\rho \\ M_{12}^{(08)}
&=& M_{21}^{(08)} = \pm \left[\left({4\over 3} m_{K^{*}} - {1\over 3} m_\rho -
m_\omega\right)\right. \\ && \times \left.\left(m_\phi - {4\over3} m_{K^{*}} +
{1\over 3} m_\rho\right)\right]^{1/2} \nonumber
\end{eqnarray}
\end{mathletters}
The eigenstates of $M^{(08)}$ are parametrized by a mixing angle $\Theta_V$
\begin{mathletters}
\label{eq:18}
\begin{eqnarray}
|\phi\rangle &=& \sin \Theta_V|\phi^{(0)} \rangle - \cos \Theta_V | \phi^{(8)}
\rangle, \\ |\omega \rangle &=& \cos \Theta_V|\phi^{(0)} \rangle + \sin
\Theta_V | \phi^{(8)} \rangle,
\end{eqnarray}
\end{mathletters}
where Eqs.~(\ref{eq:17}) imply the usual $SU(3)_V$ prediction for the tangent
of the mixing angle
\begin{equation}\label{eq:19}
\tan \Theta_V = \mp \sqrt{{m_\phi - {4\over3}m_{K^{*}} + {1 \over 3}m_\rho}
 \over {{4 \over 3}m_{K^{*}} - {1 \over 3} m_\rho - m_\omega}} \simeq \mp 0.76
 \ .
\end{equation}

In the large $N_c$ limit \cite{thooft,ven}, quark loops are suppressed, so that
the leading diagrams in the meson sector contain a single quark loop. As a
result, the octet and singlet mesons can be combined into a single ``nonet''
matrix
\begin{equation}\label{eq:21}
N_\mu = {\cal O}_\mu + {I\over\sqrt{3}}\, S_\mu,
\end{equation}
which enters the chiral Lagrangian. The kinetic, interaction and mass terms at
leading order in $1/N_c$ are
\begin{eqnarray}
{\cal L}_{\rm kin} &\rightarrow& - i \, {\rm Tr}\, N_\mu^\dagger \left( v \cdot
{\cal D} \right) N^\mu ,
\label{eq:22}
\end{eqnarray}
\begin{equation}\label{eq:22b}
{\cal L}_{\rm int} \rightarrow i g_2\, {\rm Tr}\, \left( \left\{N_\mu^\dagger,
N_\nu\right\} A_\lambda\right) v_\sigma \epsilon^{\mu\nu\lambda\sigma},
\end{equation}
and
\begin{equation}\label{eq:22c}
{\cal L}_{\rm mass} \rightarrow \mu\, {\rm Tr}\, N_\mu^\dagger N^\mu +
\lambda_2 \, {\rm Tr}\,\left( \left\{N_\mu^\dagger, N^\mu\right\} {\cal M}_\xi
\right) .
\end{equation}
Comparing with Eqs.~(\ref{kin})--(\ref{eq:13}), one finds that in the $N_c
\rightarrow \infty$ limit,
\begin{equation}\label{eq:15}
\Delta \mu \rightarrow 0, \ \sigma_0 \rightarrow 0, \ \sigma_8 \rightarrow 0,
\end{equation}
\begin{equation}\label{eq:20}
g_1 \rightarrow {{2g_2}\over \sqrt{3}},\qquad \lambda_1 \rightarrow {{2
\lambda_2} \over \sqrt{3}},\qquad \tan \Theta_V \rightarrow {1 \over \sqrt2},
\end{equation}
the $|\phi \rangle$ state becomes ``pure'' $|s\bar s \rangle$, and the nonet
matrix is
\begin{equation}\label{eq:23}
N_\mu = \left[
\begin{array}{ccc}
{\rho_\mu^0 \over \sqrt{2}} + {\omega_\mu \over \sqrt{2}} & \rho_\mu^+
&K_\mu^{*+}\\ \rho_\mu^- & -{\rho_\mu^0 \over \sqrt{2}} + {\omega_\mu \over
\sqrt{2}} & K_\mu^{*0}\\ K_\mu^{*-} & \overline K_\mu^{*0} & \phi_\mu
\end{array}
\right]\, .
\end{equation}

If the minus sign is chosen in Eq.~(\ref{eq:17}c), the prediction for the
mixing angle at leading order in chiral perturbation theory, Eq.~(\ref{eq:19}),
is close to its value for large $N_c$. At leading order in chiral perturbation
theory, the partial width for the Zweig
forbidden decay $\phi \rightarrow \rho \pi$ summed over all three modes is
\begin{equation}\label{eq:rhopi}
\Gamma \left( \phi \rightarrow \rho \pi \right) = {2 h^2 \left| \bbox{p}_\pi
\right|^3 \over \pi f^2},
\end{equation}
The coupling $h$, which vanishes as $N_c \rightarrow \infty$, is
\begin{equation}\label{eq:h}
h = {g_1 \over \sqrt 2} \sin \Theta_V - {g_2 \over \sqrt 3} \cos \Theta_V.
\end{equation}
The measured branching ratio gives $h\approx 0.05$, which also suggests that
the couplings are close to the $N_c \rightarrow \infty$ values.

In the nonrelativistic constituent quark model, assuming the $| \phi \rangle$
is pure $|s\bar s\rangle$, $g_1 = 2/\sqrt{3}$ and $g_2 = 1$. In the
nonrelativistic chiral quark model \cite{MG}, $g_1$ and $g_2$ are reduced by a
factor of 0.75 from their values in the non-relativistic constituent quark
model.

In chiral perturbation theory the leading corrections to the expressions for
the vector meson masses in Eqs.~(\ref{eq:16}) are of order $m_q^{3/2}$ (recall
we are treating $\Delta \mu$ as of order $m_q$) and arise from one-loop
self-energy diagrams giving
\begin{eqnarray}
\delta m_\rho &=& - {1\over 12\pi f^2} \left[g_2^2 \left({2\over 3} m_\pi^3 + 2
m_K^3 + {2 \over 3} m_\eta^3 \right)+ g_1^2 m_\pi^3\right]\nonumber \\ \delta
m_{K^{*}} &=& -{1\over 12\pi f^2} \left[ g_2^2 \left( {3\over 2} m_\pi^3 + {5
\over 3} m_K^3 + {1 \over 6} m_\eta^3\right) + g_1^2 m_K^3\right]\nonumber\\
\delta M_{11}^{(08)} &=& -{1\over 12\pi f^2} g_1^2 \left(3 m_\pi^3 +4 m_K^3 +
m_\eta^3 \right)\label{eq:25}\\ \delta M_{22}^{(08)} &=& - {1\over 12\pi f^2}
\left[ g_2^2 \left(2 m_\pi^3 + {2 \over 3} m_K^3 + {2 \over 3} m_\eta^3\right)
+ g_1^2 m_\eta^3\right]\nonumber\\ \delta M_{12}^{(08)} &=& \delta
M_{21}^{(08)} = {1\over 12\pi f^2} \sqrt{{2 \over 3}}g_1g_2 \,\left( -3 m_\pi^3
+ 2m_K^3 + m_\eta^3 \right).  \nonumber
\end{eqnarray}
The singlet-octet mixing angle $\Theta_V$ including these corrections is
\begin{equation}\label{eq:26}
\tan \Theta_V = \mp \sqrt{{m_\phi - {4\over 3} m_{K^{*}} + {1\over 3} m_\rho -
\delta m \over {4\over 3} m_{K^{*}} - {1\over 3} m_\rho - m_\omega + \delta m}}
\end{equation}
where
\begin{eqnarray}
\delta m &=&  - {4\over 3} \delta m_{K^{*}} + {1\over 3} \delta m_\rho + \delta
M_{22}^{(08)} \nonumber\\ &=& - {1\over 12\pi f^2} \left(g_1^2 + {2 \over 3}
g_2^2 \right) \left({1\over 3} m_\pi^3 - {4 \over 3} m_K^3 +m_\eta^3
\right).\label{eq:27}
\end{eqnarray}

Using the relation between $g_2$ and $g_1$ in Eq.~(\ref{eq:20}), we find that
\begin{equation}\label{eq:28}
\delta m \rightarrow - {2 g_2^2\over 12\pi f^2} \left({1 \over 3} m_\pi^3 -
{4\over 3} m_K^3 +  m_\eta^3\right) .
\end{equation}
With $g_2 = 0.75$, Eq.~(\ref{eq:28}) yields $\delta m \simeq - 4$~MeV.  The
combination of mass shifts, $\delta m$, that affects the mixing angle
$\Theta_V$ is very small even though the corrections to the individual masses
are substantial (e.g., $\delta m_\rho \simeq -300$~MeV). $\delta m$, which is
of order $1/N_c$, must transform like a $\bbox{27}$ of flavor $SU(3)$.
%The linear combination $ \left(m_\pi^2/3 -
%4m_K^2/3 + m_\eta^2\right)$ with $m^3 \rightarrow m^2$ vanishes by the
%Gell-Mann--Okubo formula for the meson masses, since one cannot make a
%$\bbox{27}$ from one power of the quark mass matrix.
The linear combination of pseudoscalar meson mass-cubed's in Eq.~(\ref{eq:28})
transforms like a $\bbox{27}$, and is numerically small.
This same linear combination occurs in the violation of the Gell-Mann--Okubo
formula for baryon masses~\cite{ej}.

For $N_c$ large, the $\phi \rightarrow \rho \pi$ decay amplitude is of order
$N_c^{-3/2}$, since the leading order $1/\sqrt{N_c}$ amplitude is forbidden by
Zweig's rule.  At leading order in chiral perturbation theory it occurs at tree
level because of order $1/N_c$ deviations from the relations $\tan \Theta_V =
1/\sqrt{2}$ and $g_1/g_2 = 2/\sqrt{3}$.  At order $m_s \ln m_s$ in the chiral
expansion the order $N_c^{-3/2}$ contribution arises from one-loop vertex and
wavefunction corrections calculated with vertices from the nonet Lagrange
density. The $\pi$ and $\eta$ loops do not contribute when one uses the nonet
Lagrangian. The resulting decay amplitude for each of the three $\rho \pi$
modes is
\begin{eqnarray}\label{eq:29}
&& {\cal A} (\phi \rightarrow \rho \pi) = {i\over f}
\epsilon^{\mu\nu\lambda\sigma} \epsilon_\mu (\phi) \epsilon_\nu^* (\rho)
p_{\pi \lambda} v_\sigma  (2\sqrt{m_\phi m_\rho})\nonumber \\ &&\qquad\left[
\sqrt 2\, h  - g_2^3 \left({m_K^2\over 8\pi^2
f^2}\right) \ln \left({ m_K^2 \over \mu^2} \right) + \ldots\right]
\end{eqnarray}
where the ellipsis denotes terms higher order in the chiral and $1/N_c$
expansions.
The terms of order $m_s$ have a dependence on the subtraction
point $\mu$ which cancels that of the logarithm in Eq.~(\ref{eq:29}).  With
$\mu = 1$~GeV, $g_2 = 0.75$, the magnitude of the term of
order $m_s \ln m_s$ in Eq.~(\ref{eq:29}) is about 1.5 times as large as the
measured $\phi \rightarrow \rho \pi$ decay amplitude.  This suggests that
either $g_2$ is smaller than the chiral quark model value or that there is a
partial cancellation between order $m_s \ln m_s$ and order $m_s$ contributions
to the decay amplitude.

The Frascati $\phi$ factory is expected to produce of order $10^{10}$ $\phi$'s,
allowing even very rare $\phi$ decay processes to be experimentally accessible.
The $\phi \rightarrow \rho \pi$ decay amplitude has a small parity-violating
$s$-wave amplitude that is induced by the weak interactions.  This amplitude
can be predicted in the combined limits of chiral $SU(2)_L \times SU(2)_R$
symmetry and large $N_c$.  In these limits the part of the weak Hamiltonian
that dominates the $s$-wave $\phi \rightarrow \rho \pi$ amplitude is due to
$Z^0$ exchange,
\begin{eqnarray}
{\cal H}_W &=& \eta {G_F\over 2\sqrt{2}}\left(1-{4 \over 3} \sin^2 \theta_W
\right) \nonumber \\ && \times \left(\bar s_\alpha \gamma_\mu s_\alpha\right)
\left[\bar u_\beta \gamma^\mu \gamma_5 u_\beta - \bar d_\beta \gamma^\mu
\gamma_5 d_\beta\right],\label{eq:300}
\end{eqnarray}
where $\eta\sim 1.56$~\cite{dsls} arises from QCD scaling between the weak
scale and low energies.  In the large $N_c$ limit, $\log \eta$ is of order
$1/N_c$ times logarithms of the form $\log M_W/\Lambda$, and we have chosen to
include corrections of this form \cite{bbg}.  The $\phi \rightarrow \rho \pi$
matrix element takes the form
\begin{eqnarray}
\langle\rho \pi |{\cal H}_W|\phi\rangle &=& \eta {G_F\over 2 \sqrt{2}}
\left(1-{4 \over 3} \sin^2 \theta_W
\right) f_\phi\, \epsilon_\mu (\phi) \nonumber\\
&& \qquad \times \langle \rho \pi| \bar u \gamma^\mu\gamma_5 u - \bar d
\gamma^\mu \gamma_5 d|0\rangle,\label{eq:30}
\end{eqnarray}
where the $\phi$ decay constant $f_\phi$ is defined by
\begin{equation}\label{eq:31}
\langle 0|\bar s \gamma_\mu s | \phi \rangle = f_\phi\ \epsilon_\mu (\phi).
\end{equation}
The measured $\phi \rightarrow e^+ e^-$ decay width implies that $f_\phi \simeq
(492\ {\rm MeV})^2$.  The left-handed isovector current $\bar u \gamma^\mu P_L
u - \bar d \gamma^\mu P_L d$ transforms as $(3_L, 1_R)$ under chiral $SU(2)_L
\times SU(2)_R$.  For matrix elements between the vacuum and a $\rho$ plus soft
pions this current is represented by the operator
\begin{equation}\label{eq:32}
(\bar u \gamma^\mu P_L u - \bar d \gamma^\mu P_L d) = {f_\rho\over 2 \sqrt{2
m_\rho}}\, {\rm Tr}\ (\xi {\cal O}_\mu^\dagger \xi^\dagger \tau^3),
\end{equation}
where $f_\rho$ is defined analogously to $f_\phi$, and has the value $f_\rho
\simeq (407\ {\rm MeV})^2$ from the $\rho \rightarrow e^+ e^-$ partial width.
In Eq.~(\ref{eq:32}), $\xi$ and ${\cal O}_\mu$ are the $2\times 2$ matrix
analogues of the corresponding $3\times 3$ matrices used in the case of chiral
$SU(3)_L \times SU (3)_R$.  The right-handed isovector current is given by
exchanging $\xi$ and $\xi^\dagger$ in Eq.~(\ref{eq:32}).  Using
Eq.~(\ref{eq:32}) to evaluate the matrix element in Eq.~(\ref{eq:30}) we find
that in the combined chiral and large $N_c$ limits
\begin{eqnarray}
&&\langle \rho^+ \pi^- |{\cal H}_W| \phi \rangle_{s-\rm wave} = - \langle\rho^-
\pi^+ |{\cal H}_W| \phi \rangle_{s-\rm wave} \nonumber\\ &&= -i\eta {G_F\over
\sqrt{2}}\left(1-{4 \over 3} \sin^2 \theta_W \right) \left({f_\phi f_\rho\over
f}\right) \epsilon^* (\rho) \cdot \epsilon(\phi)\label{eq:33}
\end{eqnarray}
and
\begin{equation}\label{eq:34}
\langle\rho^0 \pi^0 | {\cal H}_W| \phi\rangle_{s-\rm wave} = 0.
\end{equation}
Interference between the $s$-wave and $p$-wave amplitudes is possible for
aligned $\phi$'s but it requires a final state interaction phase.
The $s$-wave $\rho^+\pi^-$ branching ratio is $10^{-11}$ which is too small
to be measured at the Frascati $\phi$
factory. However, an enhancement of the parity-violating decay rate could make
the signal observable. This provides a very interesting test of new physics,
because it probes non-leptonic neutral currents involving strange quarks. We
will consider the application of chiral perturbation theory to other processes
such as $\phi \rightarrow \rho \gamma \gamma$ elsewhere.

This work was supported in part by the Department of Energy under Grant
Nos.~DOE-FG03-90ER40546 and DE-FG03-92-ER40701. E.J.\ was supported in part by
the NYI program, through Grant No.~PHY-9457911 from the National Science
Foundation, and by the Alfred P.~Sloan Foundation. A.M.\ was supported in part
by the PYI program, through Grant No.~PHY-8958081 from the National Science
Foundation.

\end{document}